\documentclass{emulateapj}
\usepackage{amsmath,amssymb}
\usepackage{graphicx,mathptm} 
\usepackage{times} 
\usepackage{natbib}
\def\ee{\end{equation}}
\def\be{\begin{equation}}
\def\bdm{\begin{displaymath}}
\def\edm{\end{displaymath}}

\def\jn2{J_n^2(z)}

\def\vpa{v_{\parallel }}

\def\ppa{p_{\parallel }}
\def\pper{p_{\perp }}

\def\omm{\omega }

\def\r{\right}
\def\a{\alpha }
\def\om{\omega }

\def\we{\omega _{p,e}}

\def\2kua{\sqrt{2}ku_{a,\parallel}}

\def\a{\alpha }

\def\mi2{\mu ^{-1/2}}
\def\m32{\mu ^{3/2}}
\def\ea{\epsilon _a}

\def\bet{\beta _e}

\def\ep{\epsilon}
\def\bea{\begin{eqnarray}}
\def\eea{\end{eqnarray}}
\def\b1{\beta _1}
\def\bd{\begin{displaymath}}
\def\ed{\end{displaymath}}
\def\ba{\begin{array}}
\def\ea{\end{array}}
\def\ebe{\edm\be}
\def\eba{\edm\bdm}

\def\at{\hbox{arctan}\, }

\def\ach{\hbox{arcosh}\, }
\def\pa{\partial}
\begin{document}
\shorttitle{Reactive versus kinetic electrostatic instability}
\shortauthors{Schlickeiser, Krakau \& Supsar}
\title{Plasma effects on fast pair beams II. Reactive versus kinetic instability of parallel electrostatic waves}
\author{R. Schlickeiser$^{1,2}$, S. Krakau$^{1}$, M. Supsar$^{1}$}
\affil{1 Institut f\"ur Theoretische Physik, Lehrstuhl IV:
Weltraum- und Astrophysik, Ruhr-Universit\"at Bochum, D-44780 Bochum, Germany\\
2 Research Department Plasmas with Complex Interactions, Ruhr-Universit\"at Bochum, D-44780 Bochum, Germany}
\email{rsch@tp4.rub.de, steffen.krakau@rub.de, markus.supsar@rub.de}
\begin{abstract}The interaction of TeV gamma rays from distant blazars with the extragalactic background light produces relativistic electron-positron pair beams by the photon-photon annihilation process. Using the linear instability analysis in the kinetic limit, which properly accounts for the longitudinal and the small but finite perpendicular momentum spread in the pair momentum distribution function, the growth rate of parallel propagating electrostatic oscillations in the intergalactic medium is calculated. Contrary to the claims of Miniati and Elyiv (2013) we find that neither the longitudinal nor the perpendicular spread 
in the relativistic pair distribution function do significantly affect the electrostatic growth rates. The maximum kinetic growth rate for no perpendicular spread is even about an order of magnitude greater than the corresponding reactive maximum growth rate. The reduction factors to the maximum growth rate due to the finite perpendicular spread in the pair distribution function are tiny, and always less than $10^{-4}$. We confirm the earlier conclusions by Broderick et al. (2012) and us, that the created pair beam distribution function is quickly unstable in the unmagnetized intergalactic medium. Therefore, there is no need to require the existence of small intergalactic magnetic fields to scatter the produced pairs, so that the explanation (made by several authors) of the FERMI non-detection of the inverse Compton scattered GeV gamma rays by a finite deflecting intergalactic magnetic field is not necessary. In particular, the various derived lower bounds for the intergalactic magnetic fields are invalid due to the pair beam instability argument. 
\end{abstract}
%
%
\section{Introduction}
The new generation of air Cherenkov TeV $\gamma$-ray telescopes (HESS, MAGIC, VERITAS) have detected about 30 cosmological blazars with strong TeV photon emission: the most distant ones are 3C279 (redshift $z_r=0.536$), 3C66A ($z_r=0.444$) and PKS 1510-089 ($z_r=0.361$). Any of these more distant than $z_r=0.16$ produces energetic $e^{\pm }$ particle beams in double photon collisions with the extragalactic background light (EBL). These pairs with typical Lorentz factors $\gamma =10^6\Gamma _6$ are expected to inverse Compton (IC) scatter on the cosmic
microwave background (CMB) radiation, on a typical length scale $l_{IC}\sim 0.75\Gamma _6^{-1}$ Mpc, thus producing gamma-rays with energy of order 100 GeV, which have not been detected by the FERMI satellite. Given the still relatively short distance $l_{IC}$, both pair production and IC emission occur primarily in cosmic voids of the intergalactic medium (IGM), which fill most of cosmic volume. It has been argued that the inverse Compton scattered gamma-rays then are still energetic enough for further pair-production interactions giving rise to a full electromagnetic cascade as in vacuum.

However, the pair-beam is subject to two-stream-like instabilities of
both electrostatic and electromagnetic nature (Broderick et al. 2012, Schlickeiser et al. 2012a). In this case the electromagnetic pair  cascade does not contribute to the multi-GeV flux, as most of the pair beam energy is transferred to the IGM with important consequences for its thermal history. Moreover, there is no need to require the existence of small intergalactic magnetic fields to scatter the produced pairs, so that the explanation of the FERMI non-detection of the inverse Compton scattered GeV gamma rays by a finite deflecting intergalactic magnetic field (Neronov and Vovk 2010, Tavecchio et al. 2011, Dolag et al. 2011, Taylor et al. 2012, Dermer et al. 2011, Takahashi et al. 2012, Vovk et al. 2012) is not necessary. 

In their instability analysis Schlickeiser et al. (2012a -- hereafter referred to as paper I) and Broderick et al. (2012) have approximated the pair parallel momentum distribution function $g(x)=\delta (x-x_c)$ by a sharp delta-function, where $x=\ppa /(m_ec)$ denotes the parallel pair momentum $\ppa $ in units of $m_ec=5.11\cdot 10^5$ eV/$c$ ($c$: speed of light), which is commonly referred to as reactive linear instability analysis. This approximation has been recently criticized by Miniati and Elyiv (2013), who noted that the finite momentum spread of the pair distribution function (referred to as kinetic instability study) will significantly reduce the maximum electrostatic growth rate to a level that the full electromagnetic pair cascade as in vacuum is not modified. The study of Cairns (1989), based on nonrelativistic kinetic plasma equations, indicated that the kinetic/reactive instability character depends strongly on the plasma beam and plasma background parameters, such as beam density $n_b$, beam speed $\beta _1c$ and background particle density $N_e$ and temperature $T_e$. Severe differences between reactive and kinetic instability rates occur particularly for beam to background particle density ratios exceeding $n_b/N_e>10^{-5}$. However, as argued below, in our case of pair beams in the IGM medium this ratio is of order $n_b/N_e\simeq 10^{-15}$, much below the critical value $10^{-5}$, so that we are in a regime where reactive and kinetic instability studies should not differ significantly according to Cairns (1989). 

However, as noted the work of Cairns (1989) is based on nonrelativistic kinetic plasma equations. 
It is the purpose of this work to investigate the claim of Miniati and Elyiv (2013) for parallel propagating electrostatic fluctuations using the correct relativistic kinetic plasma equations. Relativistic kinetic instability studies are notoriously difficult and complicated due to plasma particle velocities close to the speed of light. Therefore extreme care is necessary in order to include all relevant relativistic effects. We therefore will repeat in detail the linear instability analysis in the kinetic limit using the realistic pair momentum distribution function. For mathematical simplicity we will restrict our analysis to parallel wave vector orientations with respect to the direction of the TeV gamma rays generating the relativistic pairs. In our analysis we will also use a more realistic modelling of the fully-ionized IGM plasma as isotropic thermal distributions.
\section{Distribution functions and earlier reactive instability results}
\subsection{Intergalactic medium}
The unmagnetized IGM consists of protons and electrons of density $N_e=10^{-7}N_7$ cm$^{-3}$. Any neutral atoms or molecules do not participate in the electromagnetic interaction with the pairs. In paper I we have modelled the IGM plasma with the cold isotropic particle distribution functions ($a=e,p$)

\be
F_a(\ppa ,\pper )={N_e\over 2\pi \pper }\delta (\ppa )\delta (\pper ),
\label{a1}
\ee
where $\ppa $ and $\pper $ denote the momentum components parallel and perpendicular to the incoming $\gamma $-ray direction in the photon-photon collisions, respectively. Here we take into account the finite temperature $T_a$ of the IGM plasma particles, adopting the isotropic Maxwellian distribution function  

\be
F_a(p)={N_e\mu _a\over 4\pi (m_ac)^3K_2(\mu _a)}e^{-\mu _a\sqrt{1+{p^2\over m_a^2c^2}}}
\label{a2}
\ee
with $p=\sqrt{\ppa ^2+\pper ^2}$ and $\mu _a=m_ac^2/(k_BT_a)=2/\beta _a^2$, where $\beta _a=\sqrt{2k_bT_a/(m_ac^2)}$ is the thermal IGM velocity in units of the speed of light. Photoionization models of the IGM (Hui and Gnedin 1997, Hui and Haiman 2003) indicate nonrelativistic electron temperatures $T_e=10^4T_4$ K, implying very small values of $\beta _e=1.8\cdot 10^{-3}T_4^{1/2}\ll 1$ and large values of $\mu _e\gg 1$. If we scale the proton temperature $T_p=\chi T_e$, we obtain $\beta _p=\sqrt{\chi \xi }\beta _e$ with the electron-proton mass ratio $\xi =m_e/m_p=1/1836$. For proton to electron temperature ratios $\chi \ll \xi ^{-1}=1836$ we find 
that $\beta _p\ll \beta _e$.
\subsection{Intergalactic pairs from photon-photon annihilation}
Schlickeiser et al. (2012b) analytically calculated the pair production spectrum from a power law distribution of the gamma-ray beam up to the maximum energy $M$ (all energies in units of $m_ec^2$), interacting with the isotropically soft photon Wien differential energy distribution $N(k_0)\propto k_0^2\exp (-k_0/\Theta )$  representing the EBL with $\Theta \simeq 2\cdot 10^{-7}$ corresponding to 0.1 eV. They found that the pair production spectrum is highly beamed into the direction of the initial gamma-ray photons, so that a highly anisotropic, ultrarelativistic velocity distribution of the pairs results. With respect to the parallel momentum $x=\ppa /(m_ec)$ the pair momentum distribution function is strongly peaked at $M_c=\Theta ^{-1}$ for the case of effective pair production $M\gg M_c$.  The differential parallel momentum spectrum of the generated pairs can be well approximated as 

\be
n(x)=A_1e^{-{x_c\over x}}{x^{{1\over 2}-p}\over [1+({x\over x_b})^{3/2}]}H(x)
\label{a3}
\ee
with the step function $H(x)=[1+(x/|x|)]/2$, and the two characteristic normalized momenta 

\be
x_c={M_c\over \ln \tau _0},\;\;\; x_b=M_c{\tau ^{2/3}_0\over 2^{7/3}}=0.2M_c\tau ^{2/3}_0
\label{a4}
\ee
where $\tau _0=\sigma _TN_0R$, with the total number density of EBL photons $N_0\simeq 1$ cm$^{-3}$, denotes the traversed optical depth of 
gamma rays. Both characteristic momenta $x_b>x_c\gg 1$ are very large compared to unity as $M_c\simeq 2\cdot 10^6$. As noted in Schlickeiser et al. (2012b) the analytical approximation (\ref{a3}) agrees rather well with 
the numerically calculated production spectrum using the code of Elyiv et al. (2009). The parallel momentum spectrum of pairs (\ref{a3}) exhibits a strong peak at $x_c$, is exponentially reduced $\propto \exp (-x_c/x)$ at smaller momenta, and exhibits a broken power law at higher momenta (see Fig. 7 in Schlickeiser et al. 2012b). 

During this analysis here we will simplify the parallel momentum spectrum (\ref{a3}) slightly to the form 

\be
n(x)=A_0g(x),\;\; \; g(x)=x^{-s}e^{-{x_c\over x}}H(x),
\label{a5}
\ee
where we keep the essential features of the spectrum (\ref{a3}), namely the exponential reduction below $x_c$, and the power-law behavior at high parallel momentum values. But instead of allowing for the broken power-law behavior above and below $x_b$, we represent this part only as a single power law with spectral index $s=p-(1/2)$. As we will see later, this simplification only affects the damping rate of plasma fluctuations, whereas the growth rate is caused by the exponential reduction below $x_c$.
 
The associated pair phase space density is then given by 

\be
f_b(\pper ,x)={n_b\over 2\pi \pper m_ec}A_0g(x)G(\pper ,b)
\label{a6}
\ee
with the normalization factor $A_0$ determined by the total beam density 

\be
n_b=10^{-22}n_{22}=\int d^3pf_b\;\;\; \hbox{cm}^{-3}
\label{a7}
\ee
In paper I we have ignored any finite spread of the pair distribution function in perpendicular momentum $\pper $, i.e. 

\be
G(\pper )=\delta (\pper )
\label{a8}
\ee
Here we will allow for such a perpendicular spread by adopting 

\be
G(\pper ,b)={H[bm_ec-\pper ]\over bm_ec}
\label{a9}
\ee
with finite values of $b$. The special form (\ref{a9}) of the perpendicular momentum distribution function is chosen because of the limit

\be
\lim_{b\to 0} G(\pper ,b)=\delta (\pper ),
\label{a10}
\ee
which can be readily proven by inspecting with an arbitrary function $W(\pper )$ the expression

\bdm
Y=\lim_{b\to 0}\int_0^\infty d\pper \, W(\pper )G(\pper ,b)
\ebe
=\lim_{b\to 0} {1\over bm_ec}\int _0^{bm_ec}d\pper \, W(\pper )
\label{a11}
\ee
Using the Taylor expansion of the function $W$ near $\pper =0$ 

\be
W(\pper )\simeq W(\pper =0)+\pper [{dW(\pper )\over d\pper }]_{\pper =0} +\ldots 
\label{a12}
\ee
readily yields 

\bdm
Y=\lim_{b\to 0} \left[W(\pper =0)+{m_ecb\over 2}[{dW(\pper )\over d\pper }]_{\pper =0}+\ldots\r]
\ebe
=W(\pper =0)
\label{a13}
\ee
Therefore, in the limit $b=0$ the broadened perpendicular distribution function (\ref{a9}) reduces to the distribution function (\ref{a8}) with no perpendicular spread. 

Using the phase space density (\ref{a6}) with Eqs. (\ref{a5}) and (\ref{a9}) in the normalization condition (\ref{a7}) then yields 

\bdm
1=A_0\int_0^\infty dx\, g(x)
\ebe
=A_0\Gamma (s-1)U(s-1,s, x_c)\simeq A_0\Gamma (s-1)x_c^{1-s}
\label{a14}
\ee
where $\Gamma (a)$ is the gamma function and $U(a,b,z)$ denotes the confluent hypergeometric function of the second kind. Its argument $x_c$
is very large, so that we have approximated $U(s-1,s, x_c)\simeq x_c^{1-s}$ for values of $s>1$. Therefore the normalization factor has to be 

\be
A_0={x_c^{s-1}\over \Gamma (s-1)}
\label{a15}
\ee
Now we estimate the value of the maximum normalized perpendicular momentum $b$. With extensive Monte Carlo simulations Miniati and Elyiv (2013) determined the maximum angular spread of the beamed pairs to $\Delta \phi =10^{-5}$ in agreement with the kinematic estimate (see Eq. (5) of Miniati and Elyiv (2013))

\be
10^{-5}=\Delta \phi ={m_ec^2\sqrt{s_0(s_0-1)}\over 2E_{\gamma }}
<{m_ec^2s_0\over 2E_{\gamma }}={\Theta \over 2},
\label{s16}
\ee
where we use the invariant maximum center of mass energy square $s_0=E_{\gamma }\Theta /m_ec^2$. This maximum angular spread determines

\be
{p_{\perp ,\rm{max}}\over \ppa }={b\over x}=\tan \left(\Delta \phi \right)=\tan (\Theta /2)\simeq {\Theta \over 2},
\label{a17}
\ee
so that with Eq. (\ref{a4}) 

\be
b={x\Theta \over 2}\simeq {x_c\Theta \over 2}={1\over 2\ln \tau _0}={7.2\cdot 10^{-2}\over 1+{\ln \tau _3\over 3\ln 10}},
\label{a18}
\ee
which for $\tau _0=10^3\tau _3$ is well below unity. The maximum perpendicular momenta of the generated pair distribution are less than 
40 keV/c.
\subsection{Reactive instability results}
As noted before, in paper I we approximated the parallel pair distribution function (\ref{a11}) by a sharp 
delta-function $m_ecg(x)=\delta (x-x_c)$ and ignored any finite spread i.e. $G(\pper )=\delta (\pper )$. Moreover, we modelled the unmagnetized IGM as a fully-ionized cold electron-proton plasma. In agreement with the earlier reactive instability study of Broderick et al. (2012), we found that very quickly oblique (at propagation angle $\theta $) electrostatic fluctuations are excited. The growth rate $(\Im \om )_{\rm max}$ and the real part of the frequency $(\Re \om )_{\rm max}$ at maximum growth are given by 

\bdm
(\Im \om )_{\rm max}\simeq {3^{1/2}\over 2}\we \alpha (\theta )
\ebe
=1.5\cdot 10^{-6}N_7^{1/6}n_{22}^{1/3}x_{c,6}^{-1/3}\left[1-\beta _1^2\cos ^2\theta \r]^{1/3}\;\; \hbox{Hz}
\label{r1}
\ee
and 

\bdm
(\Re \om )_{\rm max}\simeq \we (1-{\a (\theta )\over 2})=\we \Bigl[1 
\edm
\be
-5\cdot 10^{-8}\left({n_{22}\over N_7x_{c,6}}\right)^{1/3}\left[1-\beta _1^2\cos ^2\theta \r]^{1/3}\Bigr],
\label{r2}
\ee
respectively, with the electron plasma frequency $\we =17.8N_7^{1/2}$ Hz. Note that we have corrected a mistake in paper I in the numerical factor in the growth rate (\ref{a12}). $n_b=10^{-22}n_{22}$ cm$^{-3}$ represent typical pair densities in cosmic voids, $x_c=10^6x_{c,6}$ 
and

\be
\a (\theta )=10^{-7}{(1-\beta _1^2\cos ^2\theta )^{1/3}n_{22}^{1/3}\over N_7^{1/3}x_{c,6}^{1/3}}\ll 1 
\label{r3}
\ee
with $\beta _1=x_c/\sqrt{1+x_c^2}$. 

The maximum growth rate occurs at the oblique angle $\theta _E=39.2$ degrees and provides as shortest electrostatic growth time 

\be
\tau _e^{-1}=\gamma _{E,\rm max}=1.1\cdot 10^{-6}{n^{1/3}_{22}N_7^{1/6}\over x^{1/3}_{c,6}}\, \hbox{Hz},
\label{r4}
\ee
Even, if nonlinear plasma effects are taken into account, we concluded in paper I that most of the 
pair beam energy is dissipated generating  electrostatic plasma turbulence, which prevents the development of a full electromagnetic pair cascade as in vacuum. 

For later comparison we note that for parallel wave vector orientations $\theta =0$ Eq. (\ref{a14}) reduce to 

\be
\a (0)=10^{-11}{n_{22}^{1/3}\over N_7^{1/3}x_{c,6}},
\label{r5}
\ee
implying for the real and imaginary frequency parts at maximum growth 
(\ref{r1}) -- (\ref{r2}) 

\bdm
(\Re \om )_{\rm max}(\theta =0)\simeq \we (1-{\a (0)\over 2})
\ebe
=\we \left[1-5\cdot 10^{-12}\left({n_{22}\over N_7x^3_{c,6}}\right)^{1/3}\r]\simeq \we
\label{r6}
\ee
and 

\be
(\Im \om )_{\rm max}\simeq {3^{1/2}\over 2}\we \alpha (0)
=1.5\cdot 10^{-10}N_7^{1/6}n_{22}^{1/3}x_{c,6}^{-1}\;\; \hbox{Hz}
\label{r7}
\ee
\section{Electrostatic dispersion relation}
The dispersion relation of weakly damped or amplified ($|\gamma |\ll \omega _R$) parallel electrostatic fluctuations with wavenumber $k$ and freuency $\omega =\omega _R+\imath \gamma $ in an unmagnetized plasma with gyrotropic distribution functions is given by (Schlickeiser 2010)

\bdm
0=\Lambda (\omega ,k)=
\ebe
1+\sum_a\frac{2\pi\omega_{p,a}^2}{\omega n_a}\int_{-\infty}^{\infty}d p_{\parallel}\, p_{\parallel}\int_0^{\infty}dp_\perp \frac{p_\perp}{\Gamma_a(\omega -k\vpa)}\frac{\partial f_a}{\partial p_\parallel}
\label{b1}
\ee
The dispersion function $\Lambda (k,\omega )$ is symmetric $\Lambda (\omega ,-k)=\Lambda (\omega ,k)$ with respect to the wavenumber $k$, so that it suffices to discuss positive values of $k>0$. Inserting the distribution functions (\ref{a2}), (\ref{a6}) and (\ref{a9}), using nonrelativistic values of $\beta _a\ll 1$, then provides

\bdm
0=\Lambda (R,I)=1-{2\we ^2n_b\over N_e}{A_0\over k^2c^2}\lim _{I\to 0} D_p(R,I,b)
\ebe
-\sum _a{\omega _{p,a}^2\over k^2c^2\beta _a^2}Z^{'}\left({z\over \beta _a}\right),
\label{b2}
\ee
where $Z^{'}(t)$ denotes the first derivative of the plasma dispersion function (Fried and Conte (1961); Schlickeiser and Yoon (2012, Appendix A)) with complex argument as $z=\omega /(kc)=R+\imath I$ with  $R=\omega _R/(kc)$ and $I=\gamma /(kc)$. For weakly damped/amplified fluctuations we use the approximations 

\bdm 
Z^{'}(t)\simeq -2\imath \pi ^{1/2}te^{-t^2}H[1-|R|]\, -2(1-2t^2),\;\;\; \hbox{for}\; |t|\ll 1,
\ebe
Z^{'}(t)\simeq -2\imath \pi ^{1/2}te^{-t^2}H[1-|R|]\, +{1\over t^2}[1+{3\over 2t^2}],\;\;\; \hbox{for}\; |t|\gg 1
\label{b3}
\ee
We notice that the imaginary part is the same in both approximations. The expression 

\bdm
D_p(R,I,b)={1\over bz}\int_0^bdq\, \int_0^\infty dx\, {x{dg(x)\over dx}\over x-z\sqrt{1+q^2+x^2}}
\eba
={1\over z}\int_0^\infty dx\, {dg(x)\over dx}
\ebe
+{1\over b}\int_0^bdq\, \int_0^\infty dx\, {{dg(x)\over dx}\over {x\over \sqrt{1+q^2+x^2}}-z},
\label{b4}
\ee
with $q=\pper /(m_ec)$, represents the pair beam contribution to the electrostatic dispersion relation. 

The first $x$-integral in Eq. (\ref{b4}) vanishes because $g(0)=g(\infty )=0$ leaving 

\be
D_p(R,I,b)={1\over b}\int_0^bdq\, \int_0^\infty dx\, {{dg(x)\over dx}\over {x\over \sqrt{1+q^2+x^2}}-R-\imath I}
\label{b5}
\ee
With Dirac's formula 

\be
\lim _{I\to 0}{1\over a-\imath I}={\cal P}{1\over a}+\imath \pi \delta (a),
\label{b6}
\ee
where ${\cal P}$ denotes the principal value, we obtain for the limit 

\bdm
\lim _{I\to 0}D_p(R,I,b)={1\over b}\int_0^bdq\, {\cal P}\int_0^\infty dx\, {{dg(x)\over dx}\over {x\over \sqrt{1+q^2+x^2}}-R}
\eba
+{\imath \pi \over b}\int_0^bdq\, \int_0^\infty dx\, {dg(x)\over dx}\delta \left({x\over \sqrt{1+q^2+x^2}}-R\right)
\eba
={1\over b}\int_0^bdq\, {\cal P}\int_0^\infty dx\, {{dg(x)\over dx}\over {x\over \sqrt{1+q^2+x^2}}-R}
\ebe
+{\imath \pi \over b}\int_0^b{dq\over 1+q^2}\, \int_0^\infty dx\, (1+q^2+x^2)^{3/2}{dg(x)\over dx}\delta \left(x-x_0(R,q)\right)
\label{b7}
\ee
with 

\be
x_0(R,q)=K(R)\sqrt{1+q^2},\;\; \; K(R)={|R|\over \sqrt{1-R^2}}
\label{b8}
\ee
The last integral has a nonvanishing value provided that $x_0(R,q)\in [0,\infty ]$, which requires subluminal real phase speed ($|R|\le 1$). 

Because of the small factor $(2n_b/N_e)\ll 1$ we ignore the contribution of the real principal part of Eq. (\ref{b7}) 
to the dispersion relation (\ref{b2}), but keep the imaginary part with the result 

\bdm
0=\Lambda (R,I)\simeq 1-\sum _a{\omega _{p,a}^2\over k^2c^2\beta _a^2}Z^{'}\left({R+\imath I\over \beta _a}\right)
\eba
-\imath {2\pi \we ^2n_bA_0\over N_ek^2c^2(1-R^2)^{3/2}b}H[1-|R|]\int_0^bdq\, \sqrt{1+q^2}[{dg(x)\over dx}]_{x_0(R,q)}
\eba
=1-{1\over \kappa ^2\beta _e^2}\left[Z^{'}\left({R+\imath I\over \beta _e}\right)+{1\over \chi }Z^{'}\left({R+\imath I\over \sqrt{\chi \xi }\beta _e}\right)\r]
\ebe
-\imath {2\pi n_b\over N_e}H[1-|R|]{x_c^{s-1}\over \kappa ^2\Gamma (s-1)(1-R^2)^{3/2}}J(b),
\label{b9}
\ee
where we have introduced the integral 

\be
J(b)={1\over b}\int_0^bdq\, \sqrt{1+q^2}[{dg(x)\over dx}]_{x_0(R,q)},
\label{b10}
\ee
the normalized wavenumber 

\be
\kappa ={kc\over \we }
\label{b11}
\ee
and the normalization constant (\ref{a15}). 

Separating the dispersion function into real and imaginary parts $\Lambda =\Re \Lambda +\imath \Im \Lambda $ we find  

\be
\Re \Lambda (R,I)=1-{1\over \kappa ^2\beta _e^2}\left[\Re Z^{'}\left({R+\imath I\over \beta _e}\right)+{1\over \chi }\Re Z^{'}\left({R+\imath I\over \sqrt{\chi \xi }\beta _e}\right)\r]
\label{b12}
\ee
and

\bdm
\Im \Lambda (R,I)=-{1\over \kappa ^2\beta _e^2}\left[\Im Z^{'}\left({R+\imath I\over \beta _e}\right)
+{1\over \chi }\Im Z^{'}\left({R+\imath I\over \sqrt{\chi \xi }\beta _e}\right)\r]
\ebe
-{2\pi n_b\over N_e}H[1-|R|]{x_c^{s-1}\over \kappa ^2\Gamma (s-1)(1-R^2)^{3/2}}J(b)
\label{b13}
\ee
We emphasize that the real part of the dispersion function (\ref{b12}) is symmetric in $R$, so that it suffices to discuss positive values of $R>0$.

It remains to calculate with the parallel pair beam distribution (\ref{a5}) 

\be
[{dg(x)\over dx}]_{x_0(R,q)}=x_0^{-(s+2)}e^{-{x_c\over x_0}}[x_c-sx_0],
\label{b14}
\ee
so that the integral (\ref{b10}) becomes 

\bdm
J(b)=
\ebe
{A\over bK^{s+1}(R)}\int_0^bdq\, {e^{-{A\over \sqrt{1+q^2}}}\over (1+q^2)^{s+1\over 2}}\left[1-{s\over A}\sqrt{1+q^2}\r]
\label{b15}
\ee
where we introduce 

\be
A(R)={x_c\over K(R)}={x_c\sqrt{1-R^2}\over R}
\label{b16}
\ee
With property (\ref{a13}) we obtain for no perpendicular spread 

\be
J(0)={A-s\over K^{s+1}(R)}e^{-A}
\label{b17}
\ee
In Appendix A we derive approximations of the integral (\ref{b15}), valid for values of $b\le b_0$, where 
$b_0=7.2\cdot 10^{-2}$, according to the estimate (\ref{a18}), is significantly smaller than unity. In terms of the value (\ref{b17}) at $b=0$ we obtain 

\be
J(b)\simeq J(0)B(X)
\label{b18}
\ee
with 

\be
X(b,A)=\sqrt{A\over 2}b,
\label{b19}
\ee
where the correction function 

\be
B(X)={e^{X^2}\over X}\left[F(X)+h(A,s)(F(X)-X)\r],
\label{b20}
\ee
with 

\be
h(A,s)={(s-1)A-s(s-2)\over 2A(A-s)}
\label{b21}
\ee
can be expressed in terms of Dawson's integral $F(X)$ (see definition (\ref{z8})). If the correction function (\ref{b20}) is smaller than unity, the perpendicular spread will reduce the growth rate $\gamma _0$ of fluctuations. If the correction function (\ref{b20}) is greater than unity, it will enhance the growth rate $\gamma _0$; each case compared to the case of no perpendicular spread $b=0$. 
\subsection{General kinetic instability analysis}
For weakly damped or amplified ($|\gamma |\ll \omega _R$) fluctuations the real and imaginary phase speed (or frequency)  parts of the fluctuations are given by (Schlickeiser 2002, p. 263)

\be
\Re \Lambda (R,I=0)=0
\label{ci1}
\ee
and 

\be
I={\gamma \over kc}=-{\Im \Lambda (R,I=0)\over {\pa \Re \Lambda (R,I=0)\over \pa R}},
\label{ci0}
\ee
respectively, where $R=\omega _R/(kc)=\omega _R/(\we \kappa )$. We then find that 

\be
\gamma (\kappa )=-\we \kappa {\Im \Lambda (R,I=0)\over {\pa \Re \Lambda (R,I=0)\over \pa R}}=\gamma _b(\kappa )-\gamma _L(\kappa )
\label{ci2}
\ee
is given by the difference of the growth rate $\gamma _p(\kappa )$ from the anisotropic relativistic pair distribution 
and the positively counted Landau damping rate $\gamma _L(\kappa )$ from the thermal IGM plasma with 

\be
\gamma _p(\kappa ,b)={2\pi \we n_b\over {\pa \Re \Lambda (R,I=0)\over \pa R}N_e}{H[1-R]x_c^{s-1}\over \Gamma (s-1)\kappa (1-R^2)^{3/2}}J(b)
\label{ci3}
\ee
and 

\bdm
\gamma _L(\kappa )={2\pi ^{1/2}\we RH[1-R]\over {\pa \Re \Lambda (R,\kappa )\over \pa R} \kappa \bet ^3}\left[e^{-{R^2\over \bet ^2}}
+{1\over \xi ^{1/2}\chi ^{3/2}}e^{-{R^2\over \xi \chi \bet ^2}}\r]
\ebe
\simeq 
{2\pi ^{1/2}\we RH[1-R]\over {\pa \Re \Lambda (R,\kappa )\over \pa R} \kappa \bet ^3}e^{-{R^2\over \bet ^2}}
\label{ci4}
\ee
\subsection{Electrostatic modes}
In Appendix B we show that the dispersion relation (\ref{ci1}) provides two collective electrostatic modes: Langmuir oscillations and ion sound waves. The Langmuir oscillations with the dispersion relation 

\be
R^2\simeq {1\over \kappa ^2}+{3\beta _e^2\over 2}={1+{3\over 2}\bet ^2\kappa ^2\over \kappa ^2}
=1+{1\over \kappa ^2}-{1\over \kappa ^2_L},
\label{ci5}
\ee
occur at normalized wavenumbers $\kappa _L\le \kappa \ll \bet ^{-1}$, where 

\be
\kappa _L^2=1+{3\bet ^2\over 2}>1
\label{ci6}
\ee
Eq. (\ref{ci5}) corresponds to the dispersion relation 

\be
\omega _R^2=\we^2[1+3k^2\lambda _{De}^2]
\label{ci7}
\ee
of Langmuir oscillations (see Appendix B). 

Likewise, the ion sound waves with the dispersion relation 

\be
R^2=R_2^2\simeq {{\xi \bet ^2\over 2}\over 1+{\bet ^2\kappa ^2\over 2}}
\label{ci8}
\ee
only exists for values of $\chi \ll 1$ or $T_p\ll T_e$ at wavenumbers $\kappa \ll (\chi ^{1/2}\bet )^{-1}=43/\beta _p$. 
Because there are no indications for such large differences in the proton to electron temperature in the IGM, we will not 
consider ion sound waves in the following.
\section{Kinetic instability analysis of Langmuir oscillations for no perpendicular spread}
We start with the case of no perpendicular spread $b=0$ in the relativistic pair distribution function. 
We use Eq. (\ref{b17}) to find for the growth rate (\ref{ci3}) 

\bdm
\gamma _p(\kappa ,b=0)
\ebe
={2\pi \we n_b\over {\pa \Re \Lambda (R,I=0)\over \pa R}N_e}{H[1-R]x_c^{s-1}(A-s)e^{-A}\over \Gamma (s-1)\kappa (1-R^2)^{3/2}K^{s+1}(R)},
\label{d0}
\ee
which is positive for values of $A>s$ corresponding to 

\be
R<{1\over \sqrt{1+(s/x_c)^2}}\simeq 1-{s^2\over 2x_c^2},
\label{d1}
\ee
given the very large value of $x_c$ (see Eq. (\ref{a4}). As long as $R\le 1-\ep $ with 

\be
\ep ={s^2\over 2x_c^2}={1\over 2}[s\Theta \ln \tau _0]^2<{\cal O}(10^{-12}),
\label{d2}
\ee
the pair parallel momentum distribution provides a positive growth rate $ \gamma _b$. 

At wavenumbers $\kappa _L<\kappa \ll \bet ^{-1}$ the dispersion relation (\ref{c8}) of Langmuir oscillations readily yields 

\bdm
{\pa \Re \Lambda (R,\kappa )\over \pa R}={2(1+\xi )\over \kappa ^2R^3}+{6\beta _e^2(1+\chi\xi ^2)\over \kappa ^2R^5}
\ebe
\simeq {2\over \kappa ^2R^5}\left[R^2+3\bet ^2\r]\simeq {2\over \kappa ^2R^3},
\label{d3}
\ee
because Langmuir oscillations occur at phase speeds $R\gg \bet $. Inserted into Eqs. (\ref{d0}) and (\ref{ci4}) the growth rate as a function of the variable (\ref{b16}) becomes 

\be
\gamma _p(A,b=0)=\gamma _p^0\kappa x_cC(A,s)
\label{d4}
\ee
with 

\be
C(A,s)={A^{s-2}(A-s)\over \Gamma (s-1)}e^{-A}
\label{d5}
\ee
and the constant 

\be
\gamma _p^0={\pi \we n_b\over N_e}H[1-R],
\label{d6}
\ee
whereas the Landau damping rate is 

\be
\gamma _L=\pi ^{1/2}\we \kappa H[1-R]R\left({R\over \bet }\right)^3e^{-{R^2\over \bet ^2}}
\label{d7}
\ee
The variable (\ref{b16}) as a function of the normalized wavenumber reads 

\be
A(\kappa )={x_c\over K(R)}={x_c\over \sqrt{{\kappa _L^2\kappa ^2\over \kappa ^2-\kappa _L^2}-1}},
\label{d8}
\ee
corresponding to 

\be
{\kappa \over \kappa _L}={1\over \sqrt{1-{\kappa _L^2\over 1+{x^2_c\over A^2}}}}\simeq 1+{\kappa _L^2\over 2(1+{x^2_c\over A^2})}
\label{d9}
\ee
\subsection{Growth rate}
In Fig. 1 we plot the growth rate $\gamma _p(b=0)/\we $ for the case of no angular spread $b=0$ as a function of the normalized wavenumber $\kappa $ for $x_c=10^6$, and different values of the spectral index $s=1.5,2,2.5$. Because of the large value of $x_c=10^6$, all growth rates peak in an extremely narrow range of wavenumber values. First, it can be seen that the weak amplification condition $\gamma _p\ll \omega _R\le \we $  is well satisfied at all values of $\kappa $. Secondly, the growth rate $\gamma _p$ exhibits a pronounced maximum. 
\begin{figure}[htbp]
\centering
\includegraphics[width=0.5\textwidth]{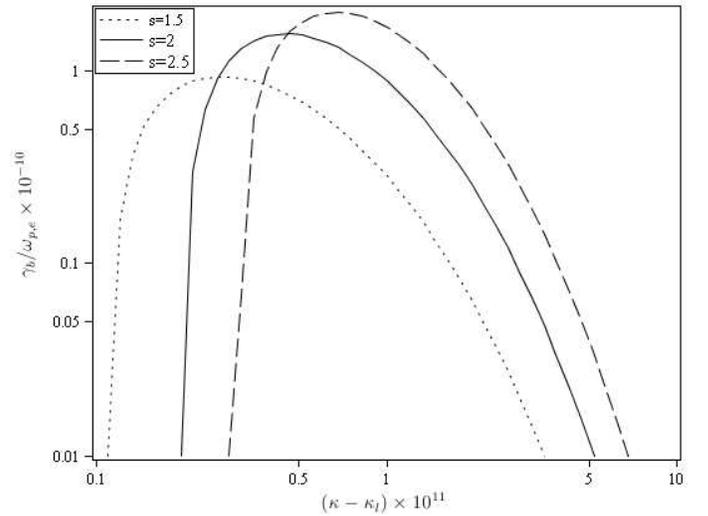}
\caption{Kinematic growth rate of parallel propagating Langmuir oscillations $\gamma _p(b=0)/\we $ for the case of no perpendicular spread ($b=0$) and the dispersion relation $\omega _R/\we $ as a function of normalized wavenumber $\kappa $ for $x_c=10^6$, $\bet =1.8\cdot 10^{-3}$ and $s=1.5,2,2.5$.}
\end{figure}
\subsection{Maximum growth rate}
\begin{figure}[htbp]
\centering
\includegraphics[width=0.5\textwidth]{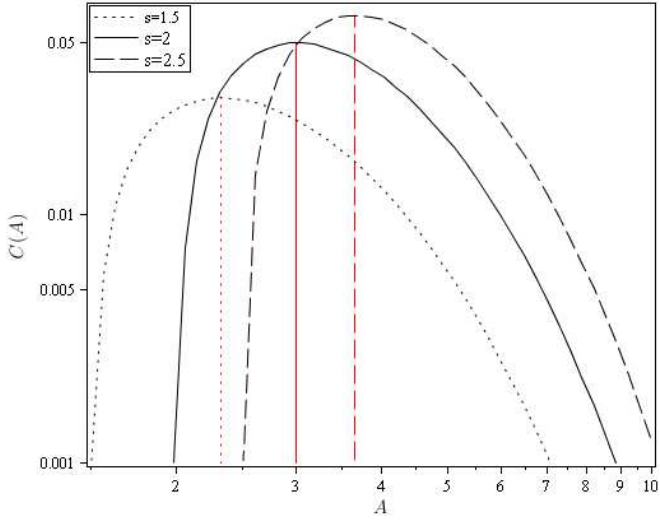}
\caption{Plot of the function $C(A)$ for three values of $s=1.5,2,2.5$ as a function of $A$.}
\end{figure}
The function $C(A,s)$, defined in Eq. (\ref{d5}), is plotted in Fig. 2 for three values of $s=1.5,2,2.5$. It has one zero at $A_N(s)=s$, is negative for smaller $A<s$, and positive for larger $A>s$, in agreement with Eq. (\ref{d1}). Extrema are located at values of $A$ satisfying 

\be
A^2-(2s-1)A+s(s-2)=0
\label{d10}
\ee
For values of $1<s\le 2$ the function $C(A,s)$ attains its maximum value at 

\be
A_0(1<s\le 2)={2s-1\over 2}\left[1+\sqrt{1+{s(2-s)\over (s-{1\over 2})^2}}\r]
\label{d11}
\ee
For the special case $s=2$ we find $A_N(2)=2$ and $A_0(2)=3$ and the maximum value 

\be
C_{\rm max}(s=2)=e^{-3}
\label{d12}
\ee
For values of $s>2$ the function $C(A,s)$ has a negative minimum at 

\be
A_{\rm min}(s>2)={2s-1\over 2}\left[1-\sqrt{1-{s(s-2)\over (s-{1\over 2})^2}}\r]
\label{d13}
\ee
and a positive maximum at

\be
A_0(s>2)={2s-1\over 2}\left[1+\sqrt{1-{s(s-2)\over (s-{1\over 2})^2}}\r]
\label{d14}
\ee
It is straightforward to show that the location of the maximum $A_0(s)<A_N(s)$ is always above the location of the zero $A_N(s)$, in 
agreement with Fig. 2. In Table 1 we calculate the locations $A_0(s)$ and values of $C_{\rm max}(s)$ for different values of $s$.

\begin{table}
\centering
\caption{Values of the zeros $A_N(s)=s$, location of maxima $A_0(s)$, maxima $C_{\rm max}(s)$ and minimum correction function $B(A_0(s),b=0.1,s)-1$ for different values of $s$ and $b=0.1$.}
\begin{tabular}{|l|l|l|l|}
\hline
$s$ & $A_0$ & $C_{\rm max}(s)$ & $B(A_0(s),b=0.1,s)-1$ \\
\hline
1.5 & 2.32 & $1.17\cdot 10^{-2}$ & $-1.35\cdot 10^{-5}$ \\
\hline
2.0 & 3.00 & $4.98\cdot 10^{-2}$ & $-2.25\cdot 10^{-5}$ \\
\hline
2.5 & 3.66 & $6.44\cdot 10^{-2}$ & $-3.35\cdot 10^{-5}$ \\
\hline
3.0 & 4.30 & $7.58\cdot 10^{-2}$ & $-4.62\cdot 10^{-5}$ \\
\hline
4.0 & 5.56 & $9.28\cdot 10^{-2}$ & $-7.62\cdot 10^{-5}$ \\
\hline
\end{tabular}
\end{table}
%
%

For ease of exposition we continue with the simplest case $s=2$. From Eq. (\ref{d4}) we then obtain for the maximum kinetic growth rate 

\be
\gamma _p^{\rm max}(b=0)={\gamma _p^0\kappa _0x_c\over e^3},
\label{d15}
\ee
which occurs at $A_0=3$, corresponding to values of $K_0(R)=x_c/3$ and values of 

\be
R^2_0={1\over 1+{9\over x_c^2}}\simeq 1-{9\over x_c^2}
\label{d16}
\ee
slightly below unity. 
\begin{figure}[htbp]
\centering
\includegraphics[width=0.5\textwidth]{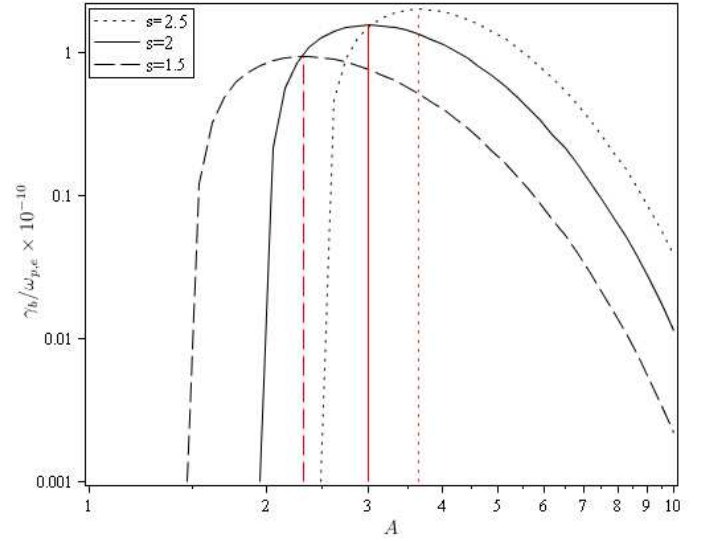}
\caption{Kinematic growth rate of parallel propagating Langmuir oscillations $\gamma _p(b=0)/\we $ for the case of no perpendicular spread ($b=0$) as a function of the variable $A$ for $x_c=10^6$, $\bet =1.8\cdot 10^{-3}$ and $s=1.5,2,2.5$.}
\end{figure}
In Fig. 3 we show the growth rate from Fig.1 now as a function of the variable $A$. We note that the location of the maximum and the zero in the case $s=2$ agree exactly with the analytical values. 

With the dispersion relation (\ref{ci5}) and the definition (\ref{ci6}) we find for the corresponding wavenumber 

\be
\kappa _0={1\over \sqrt{1-{3\bet ^2\over 2}-{9\over x_c^2}}}\simeq 1+{3\bet ^2\over 4}+{9\over 2x_c^2}\simeq 1
\label{d17}
\ee
Maximum growth occurs at frequencies

\be
\omega _{R,0}=\we \kappa _0R_0\simeq \we ,
\label{d18}
\ee
in perfect agreement with the reactive result (\ref{r6}). 

Moreover, the maximum growth rate (\ref{d15}) is given by 

\be
\gamma _p^{\rm max}(b=0)=2.8\cdot 10^{-9}{n_{22}x_{c,6}\over N_7^{1/2}}\;\; \hbox{Hz}
\label{d19}
\ee
which is about an order of magnitude larger than the maximum reactive growth rate (\ref{r7}). Apparently, the spread in parallel 
momentum of the pair distribution function does not reduce the maximum growth rate of parallel Langmuir oscillations, in disagreement with the result of Miniati and Elyiv (2013). 

At the same values of $R_0$ and $\kappa _0$, because of the exponential factor, the Landau damping rate (\ref{d7}) of Langmuir oscillations is negligibly small 

\bdm
\gamma _L(R_0)=\pi ^{1/2}\we \kappa _0R_0\left({R_0\over \bet }\right)^3e^{-{R_0^2\over \bet ^2}}
\ebe
\simeq {\pi ^{1/2}\we \over \bet ^3}e^{-{1\over \bet ^2}}<10^{-10^5}
\label{d20}
\ee
\section{Kinetic instability analysis of Langmuir oscillations for finite perpendicular spread}
With the correction function (\ref{b20}) for finite perpendicular spreads below the limit $b_0$, the growth rate in this case 

\be
\gamma _p(b)=B(X)\gamma _p(b=0)
\label{f1}
\ee
is simply related to the growth rate $\gamma _p(b=0)$. The growth rate $\gamma _p(b)$ with finite spread as compared to the growth rate $\gamma _p(b=0)$ with no finite spread is enhanced (reduced) if the correction function (\ref{b20}) is greater (smaller) than unity.  
The correction function (\ref{b20}) reads 

\be
B(X)=B(A,b,s)={e^{X^2}\over X}\left[(1+h)F(X)-hX\r]
\label{f2}
\ee
with the function 

\be
h(A,s)={(s-1)A-s(s-2)\over 2A(A-s)}
\label{f3}
\ee
We noted before that the growth rate $\gamma _p(b=0)$ is positive only for values of $A>s$, so we restrict our analysis to this range. 
For $A>s$ the function (\ref{f3}) is positive for all values of $A>s>1$. With $A=s+t$ the function (\ref{f3}) reads 

\be
h(t,s)={s+(s-1)t\over 2t(t+s)}={s-1\over 2(t+s)}+{s\over 2t(t+s)}
\label{f4}
\ee
with $t\in (0,\infty]$. The function is strictly decreasing, as 

\be
{dh(t,s)\over dt}=-{(s-1)t^2+2st+s^2\over 2t^2(t+s)^2}
\label{f5}
\ee
is always negative. No extreme values occur in the interval $(0,\infty ]$. For later use we note that the condition $h(A,s)=1/2$ leads to the equation 

\be
A^2-(2s-1)A+s(s-2)=0,
\label{f51}
\ee
which is identical to Eq. (\ref{d10}), determining the maximum growth rate $\gamma ^{\rm max}_p(b=0)$ through the function $C(A,s)$. 
Hence, at the maximum $A_0(s)$ the function 

\be
h(A_0(s),s)={1\over 2}
\label{f52}
\ee
for all values of $s$. Moreover, for larger values of $A>A_0(s)$, the function $h(A,s)<1/2$.
\subsection{Correction function for the maximum growth rate}
The maximum growth rate $\gamma _p^{\rm max}(b=0)$ occurs at $A_0(s)$ listed in Table 1. For values of $b<0.1$ the variable (\ref{b19}) 

\be
X=\sqrt{A_0(s)\over 2}b<0.071\sqrt{A_0(s)}<0.17
\label{g1}
\ee
is smaller than unity for all values of $b<0.1$, because for $s\le 4$ we calculated $A_0(s)\le A_0(4)=5.56$. We therefore use the series expansion (\ref{z11}) for Dawson's integral in Eq. (\ref{f2}) to find 

\bdm
B(X\ll 1)\simeq e^{X^2}\left[1-{2\over 3}(1+h)X^2(1-{2\over 5}X^2)\r]
\ebe
\simeq 1-{2h-1\over 3}X^2-{4h-1\over 10}X^4,
\label{g2}
\ee
With the value (\ref{f52}) the quadratic terms vanishes and we obtain the correction 

\be
B(A_0(s),b,s)\simeq 1-{X^4\over 10}=1-{A^2_0(s)b^4\over 40}
\label{g3}
\ee
For the maximum value of $b=0.1$, we calculate the reduction factor $B(A_0(s),b,s)-1$ for different values of $s$. The results are listed in Table 1. As can be seen, the reduction factors due to the finite spread in the pair distribution function are tiny, always less than $(-10^{-4})$. Contrary to the statement of Miniati and Elyiv (2013) we find that the finite perpendicular spread does not significantly reduce the maximum growth rate. 
\subsection{General behavior of the correction function}
Dawson's integral satisfies the linear differential equation 

\be
{dF(X)\over dX}=1-2XF(X),
\label{g4}
\ee
so that the first derivative of the correction function (\ref{f2}) is given by 

\be
{\pa B(X)\over \pa X}={e^{X^2}\over X^2}\left[(1+h)X-(1+h)F(X)-2hX^3\r]
\label{g5}
\ee
The extreme value of the correction function $B(X_E)$ occurs at $X_E$ given by the solution of the transcendental equation 

\be
X_E-F(X_E)={2h\over 1+h}X_E^3
\label{g6}
\ee
Inserting this condition into Eq. (\ref{f2}) we obtain for the extreme value of the correction function 

\be
B_E=e^{X_E^2}\left[1-2hX_E^2\r]
\label{g7}
\ee
We recall that for values of $A>A_0(s)$, corresponding to $X_E>b\sqrt{A_0(s)/2}$, the function $h<1/2$. The first and second derivative of function (\ref{g7}) are given by 

\be
{dB_E\over dX_E}=2X_Ee^{X_E^2}\left[(1-2h)-2hX_E^2\r]
\label{g8}
\ee
and

\be
{d^2B_E\over dX^2_E}=2e^{X_E^2}\left[(1-2h)+2(1-5h)X_E^2-4hX_E^4\r]
\label{g9}
\ee
The function $B_E$ has a single maximum at $X^2_E=(1-2h)/2h$ given by 

\be
B_E^{\rm max}=2he^{{1\over 2h}-1}
\label{g10}
\ee

For given $b$, Eq. (\ref{g7}) corresponds to the extreme value 

\bdm
B_E(A_E)=e^{b^2A_E\over 2}\left[1-hb^2A_E\r]
\ebe
=e^{b^2A_E\over 2}\left[1-{b^2\over 2}\left(s-1-{s\over A_E-s}\right)\r],
\label{g11}
\ee
where we inserted the function $h(A_E,s)$ from Eq. (\ref{f3}). Even without knowing the value $A_E$, we can draw some interesting conclusions from Eq. (\ref{g11}). 

For values of $s\ll A_E\ll (2/b^2)$ the function (\ref{g11}) approaches 

\be
B_E(s\ll A_E\ll {2\over b^2})\simeq 1-{b^2(s-1)\over 2},
\label{g12}
\ee
producing at most a tiny correction over a wide range of $s\ll A_e\ll 200$ in agreement with our earlier discussion of the maximum growth rate. 
\section{Summary and conclusions}
The interaction of TeV gamma rays from distant blazars with the extragalactic background light produces relativistic electron-positron pair beams by the photon-photon annihilation process. The created pair beam distribution is unstable to linear two-stream instabilities of both electrostatic and electromagnetic nature in the unmagnetized intergalactic medium. Based on a linear reactive instability analysis Broderick et al. (2012) and Schlickeiser et al. (2012) have concluded that the created pair beam distribution function is quickly unstable to the excitation of electrostatic oscillations in the unmagnetized intergalactic medium, so that the generation of inverse-Compton scattered GeV gamma-ray photons by the pair beam is significantly suppressed. Because most of the pair kinetic energy is transferred to electrostatic fluctuations, less kinetic pair energy is available for inverse Compton interactions with the microwave background radiation fields. 
Therefore, there is no need to require the existence of small intergalactic magnetic fields to scatter the produced pairs, so that the explanation (made by several authors) of the FERMI non-detection of the inverse Compton scattered GeV gamma rays by a finite deflecting intergalactic magnetic field is not necessary. In particular, the various derived lower bounds for the intergalactic magnetic fields are invalid due to the pair beam instability argument. 

Miniati and Elyiv (2013) have argued that the more appropriate linear kinetic instability analysis, accounting for the longitudinal and the small but finite perpendicular momentum spread in the pair momentum distribution function, significantly reduces the growth rate of electrostatic oscillations by orders of magnitude compared to the linear reactive instability analysis, concluding that the pair beam instability does not modify the pair cascade as in vacuum. We therefore have repeated the linear instability analysis in the kinetic limit for parallel propagating electrostatic oscillations using the realistic pair distribution function with longitudinal and perpendicular spread. Contrary to the claims of Miniati and Elyiv (2013) we find that neither the longitudinal nor the perpendicular spread in the relativistic pair distribution function do significantly affect the electrostatic growth rates. The maximum kinetic growth rate for no perpendicular spread is even about an order of magnitude greater than the corresponding reactive maximum growth rate. The reduction factors to the maximum growth rate due to the finite perpendicular spread in the pair distribution function are tiny, and always less than $10^{-4}$. We confirm the earlier conclusions by Broderick et al. (2012) and Schlickeiser et al. (2012a), that the created pair beam distribution function is quickly unstable in the unmagnetized intergalactic medium. 

As our analysis has shown, relativistic kinetic instability studies are notoriously difficult and complicated due to plasma particle velocities close to the speed of light. Therefore extreme care is necessary in order to include 
all relevant relativistic effects, as done in the present study.  
\begin{acknowledgements} 
We gratefully acknowledge partial support of this work by the Mercator Research Center Ruhr (MERCUR) through grant Pr-2012-0008, and the German Ministry for Education
and Research (BMBF) through Verbundforschung Astroteilchenphysik grant 05A11PCA. 
\end{acknowledgements}

\section{Appendix A: Approximations of the integral $J(b)$}
We introduce 

\be
T(b,A,s)={K^{s+1}(R)\over A}e^AbJ(b),
\label{z1}
\ee
so that according to Eqs. (\ref{b15}) 

\be
T(b,A,s)=e^{A}\int_0^bdq\, {e^{-{A\over \sqrt{1+q^2}}}\over (1+q^2)^{s+1\over 2}}\left[1-{s\over A}\sqrt{1+q^2}\r]
\label{z2}
\ee
The substitution $q=\tan t$ provides 

\bdm
T(b,A,s)=e^{A}\Bigl[\int_0^{\at b}dt\, \cos ^{s-1}t\, e^{-A\cos t} 
\eba
-{s\over A}\int_0^{\at b}dt\, \cos ^{s-2}t\, e^{-A\cos t}\Bigr]
\ebe
=Y(b,A,p={s-1\over 2})-{s\over A}Y(b,A,p={s-2\over 2})
\label{z00}
\ee
with 

\be
Y(b,A,p)=e^{A}\int_0^{\at b}dt\, \cos ^{2p}t\, e^{-A\cos t}
\label{z01}
\ee
Because $b$ is significantly smaller than unity, we approximate 

\be
\cos t\simeq 1-{t^2\over 2},
\label{z3}
\ee
so that with $\at b\simeq b$ 

\be
Y(b,A,p)\simeq \int _0^{b}dq\,[1-pq^2] e^{{Aq^2\over 2}}
\label{z4}
\ee
We restrict our analysis to values of $\bet \le R\ll R_0$, where $R_0$ denotes the real phase speed (\ref{d16}), where the maximum growth rate $\gamma ^{\rm max}_p(b=0)$ for no angular spread occurs (see Sect. 4.2). In this case the variable (\ref{b16})

\be
A(R\le R_0)\ge A(R_0)>3
\label{z6}
\ee
is always larger than $3$. The main contribution to the integral (\ref{z4}) is then indeed provided by small values of $q\ll 1$, so that the approximation (\ref{z3}) is justified. 

The integral (\ref{z4}) can be expressed in terms of Dawson's integral (Abramowitz and Stegun 1972, Ch. 7.1; Lebedev 1972, Ch. 2.3), the error function of imaginary argument, 

\be
F(x)=e^{-x^2}\int_0^xdt\, e^{t^2}
\label{z8}
\ee
as 

\be
Y(b,A,p)=\sqrt{2\over A}e^{X^2}\bigl[F\left(X\right)+{p\over A}[F\left(X\right)-X]\bigr],
\label{z9}
\ee
with 

\be
X(b,A)=\sqrt{A\over 2}b
\label{z10}
\ee
Dawson's integral (\ref{z8}) has a maximum $F_m=0.541$ at $x_m=0.924$, the series expansion 

\be
F(x)=\sum_{n=0}^\infty {(-1)^n2^nx^{2n+1}\over 1\cdot 3\cdot \cdot \cdot (2n+1)}=x\left[1-{2\over 3}x^2+{4\over 15}x^4\mp \ldots \r]
\label{z11}
\ee
and the asymptotic expansion 

\be
F(x\gg 1)\simeq {1\over 2x}[1+{1\over 2x^2}+{3\over 4x^4}]
\label{z12}
\ee
\begin{figure}[htbp]
\centering
\includegraphics[width=0.5\textwidth]{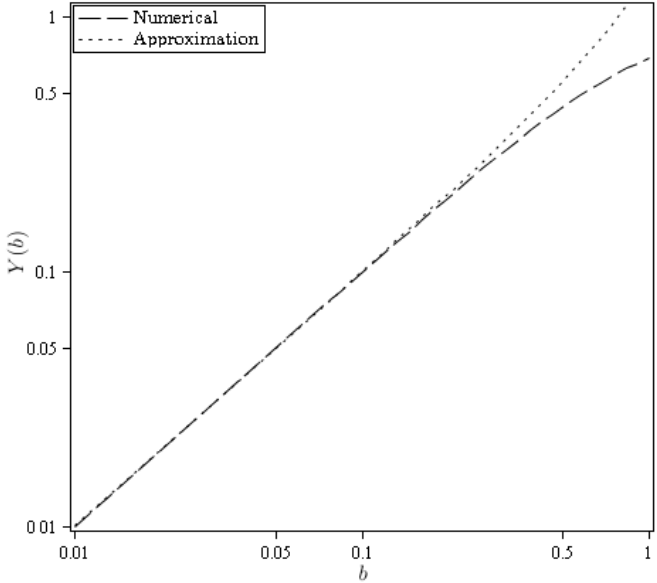}
\caption{Comparison of the numerically evaluated exact integral (\ref{z01}) with its approximation (\ref{z9}) for $p=2$ and $A=3$.}
\end{figure}
\begin{figure}[htbp]
\centering
\includegraphics[width=0.5\textwidth]{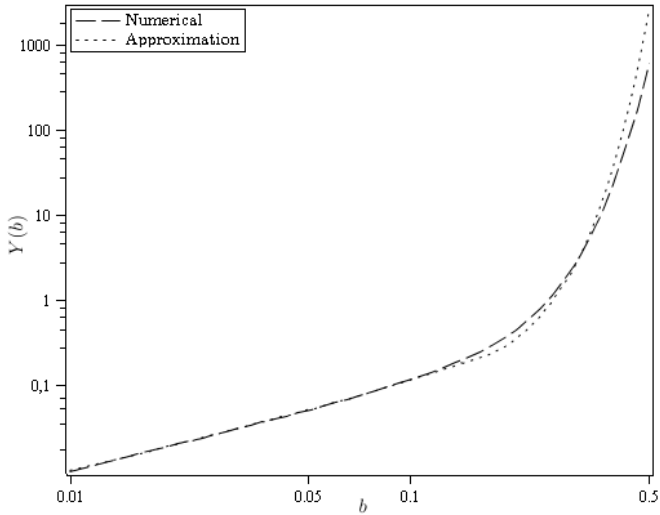}
\caption{Comparison of the numerically evaluated exact integral (\ref{z01}) with its approximation (\ref{z9}) for $p=2$ and $A=100$.}
\end{figure}
In Figures 4 and 5 we compare the numerically evaluated exact integral (\ref{z01}) with its approximation (\ref{z9}) for $p=2$ and two values of $A=3$ and $A=100$. In both cases the agreement is excellent for values of $b<0.1$.

According to Eqs. (\ref{z1}) and (\ref{z00}) we obtain the approximations 

\bdm
J(b)\simeq {Ae^{-A}\over K^{s+1}(R)}{e^{X^2}\over X}\Bigl[(1-{s\over A})F(X)
\ebe
+\left((s-1)A-s(s-2)\right){F(X)-X\over 2A^2}\Bigr]
\label{z13}
\ee
The small argument expansion (\ref{z11}) readily yields 

\be
J(0)=J(b=0)={A-s\over K^{s+1}(R)}e^{-A},
\label{z14}
\ee
which agrees with Eq. (\ref{b17}), so that the correction function (\ref{b20}) becomes 

\be
B(X)={e^{X^2}\over X}\left[F(X)+h(A,s)(F(X)-X)\r],
\label{z15}
\ee
with 

\be
h(A,s)={(s-1)A-s(s-2)\over 2A(A-s)}
\label{z16}
\ee
\section{Appendix B: Collective electrostatic modes}
Eq. (\ref{ci1}) together with the real part of the dispersion relation (\ref{b12}) reads 
 
\bdm
0=\Re \Lambda (R,I=0)=1-
\ebe
{1\over \kappa ^2\beta _e^2}\left[\Re Z^{'}\left({R\over \beta _e}\right)+{1\over \chi }\Re Z^{'}\left({R\over \sqrt{\chi \xi }\beta _e}\right)\r]
\label{c1}
\ee
In order to use the asymptotic expansions (\ref{b3}) for proton-electron temperature ratios $\chi <\xi ^{-1}=1836$ we have to consider three cases: 

(a) In the case of phase speeds larger than $\bet $, 

\be
R\gg \beta _e. 
\label{c3}
\ee
both arguments of the $Z^{'}$-function are large compared to unity, so that we may use the asymptotic expansion 

\be
\Re Z^{'}(t\gg 1)\simeq {1\over t^2}[1+{3\over 2t^2}]
\label{c4}
\ee

(b) In the case of intermediate phase speeds, 

\be
{R\over \beta _e}\ll 1\ll {R\over \sqrt{\chi \xi }\beta _e},
\label{c5}
\ee
we use the expansion (\ref{c4}) in the third term of Eq. (\ref{c1}) and the asymptotic expansion for small arguments 

\be
\Re Z^{'}(t\ll 1)\simeq -2[1-2t^2]
\label{c6}
\ee
in the second term of Eq. (\ref{c1}).

(c) In the case of very small phase speeds,

\be
R\ll \sqrt{\chi \xi }\beta _e,
\label{c7}
\ee
we use the expansion (\ref{c6}) in the second and third term of Eq. (\ref{c1}).

We consider each case in turn. 
\subsection{Large phase speed $R\gg \beta _e$}
Here we readily obtain for Eq. (\ref{c1}) 

\be
\Re \Lambda (R,\kappa )=1-{1+\xi \over \kappa ^2R^2}-{3\beta _e^2(1+\chi\xi ^2)\over 2\kappa ^2R^4}=0
\label{c8}
\ee
yielding the dispersion relation 

\be
R^4-{1+\xi \over \kappa ^2}R^2-{3\beta _e^2\over 2\kappa ^2}=0
\label{c9}
\ee
with the solution 

\be
R^2={1+\xi \over 2\kappa ^2}\left[1+\sqrt{1+{6\beta _e^2\kappa ^2\over (1+\xi )^2}}\r]\simeq 
{1\over 2\kappa ^2}\left[1+\sqrt{1+{6\beta _e^2\kappa ^2}}\r]
\label{c10}
\ee
The requirement $R\gg \beta _e$ implies the wavenumber restriction $\beta _e^2\kappa ^2\ll 2.5$. Likewise, the subluminality requirement $R<1$ demands 

\be
\kappa ^2>\kappa _L^2=1+{3\bet ^2\over 2}
\label{c101}
\ee
In this wavenumber range the solution (\ref{c9}) reduces to 

\be
R^2\simeq {1\over \kappa ^2}+{3\beta _e^2\over 2}={1+{3\over 2}\bet ^2\kappa ^2\over \kappa ^2},
\label{c11}
\ee
corresponding to Langmuir oscillations 

\be
\omega _R^2=\we^2[1+3k^2\lambda _{De}^2]
\label{c12}
\ee
for $2^{-1/2}\bet \le k\lambda _{De}\ll 1$ with the electron Debye length $\lambda _{De}=\beta _ec/\sqrt{2}\we $.  
\subsection{Intermediate phase speed $\sqrt{\chi \xi}\beta _e=\beta _p\ll R\ll \beta _e$}
In this case we derive for Eq. (\ref{c1}) 

\be
\Re \Lambda (R,\kappa )\simeq 1+{2\over \bet ^2\kappa ^2}-{\xi \over \kappa ^2R^2}-{4R^2\over \bet ^4\kappa ^2}=0,
\label{c13}
\ee
yielding the dispersion relation 

\be
R^4-{\bet ^2\over 2}(1+{\bet ^2\kappa ^2\over 2})R^2+{\xi \bet ^4\over 4}=0
\label{c14}
\ee
with the two formal solutions

\bdm
R^2_{1,2}={\bet ^2\over 4}(1+{\bet ^2\kappa ^2\over 2})\left[1\pm \sqrt{1-{4\xi \over (1+{\bet ^2\kappa ^2\over 2})^2}}\r]
\ebe
\simeq {\bet ^2\over 2}(1+{\bet ^2\kappa ^2\over 2})\left[1\pm \left(1+{\xi \over (1+{\bet ^2\kappa ^2\over 2})^2}\right)\r]
\label{c15}
\ee
The first solution 

\be
R^2_1\simeq \bet ^2(1+{\bet ^2\kappa ^2\over 2})
\label{c16}
\ee
violates the restriction $R^2\ll \bet ^2$, leaving as only solution 

\be
R^2=R_2^2\simeq {{\xi \bet ^2\over 2}\over 1+{\bet ^2\kappa ^2\over 2}}
\label{c17}
\ee
This ion sound wave solution has to fulfill the second restriction $R^2\gg \chi \xi \bet ^2$, corresponding to the condition 

\be
1+{\bet ^2\kappa ^2\over 2}\ll{1\over 2\chi },
\label{c18}
\ee
which is only possible for values of $\chi \ll 1$ or $T_p\ll T_e$. In this case the solution (\ref{c17}) holds for wavenumbers 
$\kappa ^2\bet ^2\ll \chi ^{-1}$. Therefore the ion sound wave solution only exists for $T_p\ll T_e$ at 
wavenumbers $(\lambda _{De}k)^2\ll (2\chi )^{-1}$ with frequencies 

\be
\omega _R^2={\beta _p^2c^2k^2\over 2(1+\lambda ^2_{De}k^2)},\,\;\; \lambda ^2_{De}k^2\ll {1\over 2\chi }={T_e\over 2T_p}
\label{c19}
\ee
\subsection{Very small phase speed $R\ll \sqrt{\chi \xi}\beta _e=\beta _p$}
In this case we derive for Eq. (\ref{c1}) 

\be
\Re \Lambda (R,\kappa )\simeq 1+{2(1+\chi )\over \chi \bet ^2\kappa ^2}-{4(1+\xi \chi ^2)R^2\over \xi \chi ^2\bet ^4\kappa ^2}=0,
\label{c20}
\ee

yielding the dispersion relation 

\be
R^2={(1+\chi )\chi \xi \bet ^2\over 2(1+\xi \chi ^2)}+{\xi \chi ^2\bet ^4\kappa ^2\over 4(1+\xi \chi ^2)}
\label{c21}
\ee
The very small phase speed requirement $R^2\ll \chi \xi \bet ^2$ corresponds to 

\be
{(1+\chi )\over 2(1+\xi \chi ^2)}+{\chi \bet ^2\kappa ^2\over 4(1+\xi \chi ^2)}\ll 1,
\label{c22}
\ee
which cannot be fulfilled. Therefore no electrostatic mode with very small phase speeds exists.

\end{document}